# A TAYLOR-MADE ARITHMETIC MODEL OF THE GENETIC CODE AND APPLICATIONS

Tidjani Négadi[*]

**Abstract**: *An important and notable improvement of our recent arithmetic model of the standard genetic code based on Gödel encoding is presented using a "head/tail" metaphor. Etched in the "head" we have the decimal representation of 23!, with 23 digits, from which Rumer's division of the genetic code table into the two equal moieties $M_1$ and $M_1$ arises quite naturally. By ultimately reducing every digit-value in the model to its prime-factorization form, we compute the exact degeneracy of the two sets as well as the number of amino acids and stop signals in them and deduce also the degeneracies and the number of amino acids of the five degeneracy-classes of the standard genetic code as well as the three stop signals. This is made possible thanks to a subtle encoding of the doublet-parts of the three sextets, in the doublets. In the "tail" side, on the other hand, we have (also etched) the prime-factorization of 23!, using the fundamental theorem of arithmetic, and leading directly to two other important divisions of the genetic code table into two equal moieties the first-base pyrimidine/purine series and the first-base keto/amino series. In both cases, the degeneracy of the two moieties appears (conspicuously) to be encoded in a Gödelian way, in the exponents of the prime factorization of 23!. This new understanding (two sides of the same "coin") improves and enriches our model and also paves the way for interesting applications inside and outside the genetic code. In the first instance we successfully include in our model the $21^{st}$ and $22^{nd}$ coded amino acids Selenocysteine and Pyrrolysine as an almost trivial application. In the second, we derive the carbon and hydrogen atom composition as well as the atom number in the "61 amino acids" of the whole genetic code table, according to characteristic patterns, from the digits representing the amino acids themselves and show also that the degeneracy and the chemical composition of the amino acids are connected. In particular* we establish a link, with passage formulae, between our arithmetic model and the recent classification by Račocević s called by its author the "Cyclic Invariant Periodic System" or CIPS. In the third, we show that specific ratios of certain arithmetic quantities associated to the amino acids classes could fit reasonably the experimental amino acids composition, in the corresponding classes, and averaged over a large set of recently analyzed proteomes covering the three phylogenetic domains of life. Finally, we end this paper with some closing remarks, in particular, we let our arithmetic model and its extensions constitute a bridge between two important approaches of the genetic code, the first one we call shCherbak's approach and the other we call Downes-Richardson's approach, and show that they are in fact, also, only "two sides of the same coin".

---

[*] *Address*: Department of Physics, Faculty of Science, University of Oran, Es-Sénia, 31100, Oran, Algeria.email: tnegadi@gmail.com; negadi_tidjani@univ-oran.dz

## 1. INTRODUCTION

Since the experimental decyphering of the genetic code in the sixties of the last Century, the theoretical investigations aiming at understanding its (mathematical) structure, especially its degeneracy *multiplet* structure, were centered mainly around the use of group theoretical technics, inherited in great part from particle physics and introduced by Hornos's group in the nineties (see Hornos et al., 2001, and the references therein). Now and since about the beginning of the present Century, we see that the group-theory road becomes somewhat breathless and various other interesting approaches are being born (see for example the review by Gusev and Schulze-Makuch, 2004). One of them, which has been of great inspiration to the present author, was published in 2003 by shCherbak (shCherbak, 2003) and shows clearly the existence of several arithmetic regularities inside the genetic code. Other interesting approaches describing various arithmetic regularities inside the genetic code exist as for example Rakočević's Harmonic Structures determined by using "Gauss's arithmetical algorithm" and leading ultimately to his recent Cyclic Invariant Periodic System classification (Rakočević, 2009, and the references therein) "encoding" the physicochemical properties of the twenty canonical amino acids in a tight manner. The original idea which has led to the present author's own approach germinated some two years ago after a long focus on the *number* 23 which role in the context of the genetic code was emphasized long time ago by Gavaudan (Gavaudan, 1971), with respect to the usual "magic" number 20. The sudden realization that the number of permutations (of 23 objects), i.e. 23!, had just the right (arithmetic) properties in our universal place-value Hindu-Arabic decimal system to represent the genetic code (Négadi, 2007) was instrumental for the following development of the model. More precisely, the twenty canonical amino acids, the building-blocks of all living beings, could be associated successfully to the nine decimal digits, *all* present in 23! with possible multiplicity, and importantly including zeros. Several other interesting results could be derived from this number. (It is custom in mathematical biology to assign numbers, in binary, quaternary, decimal, etc., to the nucleobases T/U, C, A and G, for example, so why-not for the amino acids?) From the group theoretical point of view it is interesting to note that the total number of *irreducible representations* of the symmetric (or permutation) group $S_{23}$, which is also equal to the total number of partitions of the integer 23, is equal to 1255. This last number coincides with the total number of *nucleons* in the 20 amino acids side-chains (see Table 2) which, in turn, is linked to the number of carbon atoms in the 20 canonical amino acids side-chains, 67, via 1255=67 (mod 99). In fact, still from the group theoretical point of view, it is even possible to put the totality of the nucleons in the 23 amino acids side-chains (1443, see at the end of the paper) and those in the five constituting atoms hydrogen, carbon, nitrogen, oxygen and sulfur (75=1+12+14+16+32) in *one* irreducible representation, corresponding to the partition $(2^3, 1^{17})$ of $S_{23}$, with dimension 1518. This point will be considered in a future paper. As a further interesting "trump" concerning the above (large) number, 23!, and coming from physics, we note that this latter written $0.2585... \times 10^{23}$, has the same magnitude as the *informational complexity* of a *conscious living creature* calculated by the physicist Freeman Dyson thirty years ago: $Q=10^{23}$ bits (Dyson, 1979). Dyson says for example in page 454 of his paper: *Q is a pure number expressing the amount of information that must be processed in order to keep the creature alive long enough to say "Cogito, ergo sum"*. Time has shown us that the choice of this number was fruitful. In this contribution to Festival Symmetry 2009 we present a novel form of our arithmetic model of the standard genetic code, using a "head/tail" picture. This is presented in section 2. Starting from the decimal representation of 23! (the "head") we show that the sorting of the 23 digits into six sub-sets of digits or "multiplets" representing the five usual multiplets of the standard genetic code and the three stop signals gives us, in a *one-step-computation*, Rumer's division of the genetic code table into its two equal moieties $M_1$ and $M_2$. We derive immediately the degeneracy of these latter. Also, using the subtle and remarkable digit-structure associated to the doublets, we infer the degeneracy of the five multiplets mentioned above according to two possible "pictures", 20 amino acids (aas) or 23 amino acids signals (AAS). On the other side (the "tail"), from which one could alternatively start, we have the prime-factorization of 23! using the fundamental theorem of arithmetic, giving the *Gödel encoding* of a number sequence S which appears to describe remarkably the degeneracy according to two other possible divisions of the genetic code table into two equal moieties, *the first-base pyrimidine/purine series* $U^{(1)}+C^{(1)}/A^{(1)}+G^{(1)}$ and the *first-base keto/amino series*, $U^{(1)}+G^{(1)}/A^{(1)}+C^{(1)}$. In this way the Gödel encoding of the sequence S gives back the decimal place-value representation of 23! and, at the same time, starting the reading from the end of sub-section 2.2 one is brought back to the beginning of sub-section 2.1. In a third section, devoted to applications, we first show that our arithmetic model is capable to include (almost trivially) the 21$^{st}$ or/and 22$^{nd}$ (*co-translationally*) coded amino acid(s) Selenocysteine and Pyrrolysine. Second, we derive the carbon and hydrogen atom composition of the amino acids as well as the atom number in the "61 amino acids" of the whole genetic code table, according to characteristic patterns, from the digits representing the amino acids themselves and show also that the degeneracy and the chemical composition of the amino acids (aas) are connected. Also, we compute the hydrogen atom number in 20 aas and in "61 aas" as well as the atom number in "61 aas", directly from the decimal representation *and* the prime factorization of 23!. We also establish a link between our arithmetic model and the Cyclic Invariant Periodic System, CIPS, of the genetic code (Račoćevic, 2009). In particular, we derive from our equations the carbon and hydrogen numbers as well as the atom numbers of the amino acids in the CIPS, as arranged according to his two *superclasses*, the *primary superclass* (PSC) and the *secondary superclass* (SSC). Other interesting mathematical derivations (from 23!), having something to do with the chemical composition mentioned above, are also reported in the second section. Finally, as another and last application in this paper, this time *outside* the genetic code itself, we show that *specific ratios* of certain arithmetic quantities associated to the amino acids classes in our model could fit favourably the experimental amino acids composition, in the corresponding classes, and averaged over a large set of recently analyzed proteomes covering the three phylogenetic domains of life (Tekaia and Yaramian, 2006). The study of amino acids composition and also the chemical (atomic) composition in proteomics are active fields of research these recent years. For example, it has been recognized that elemental (ex: nitrogen, carbon) composition of genomes and proteins can be related to resource limitation (Bragg and Hyder, 2004). The mathematical tools used in this paper are of two kinds. First, we use as custom some few arithmetic functions to reveal genetic information. Two of these functions called $a_0(n)$ and $a_1(n)$ and defined for any integer n>2. The former gives the sum of the *prime factors* of n, from its prime factorization via the fundamental theorem of arithmetic, counting multiplicities, and the latter gives also the sum of the prime factors but ignoring the multiplicities. Another one, $\Omega(n)$, called the Big Omega Function, gives the number of prime factors of n, counting the multiplicities. Second, we also use in our applications two elementary mathematical processes, or algorithms, that we have found interesting to beget genetic information too. For example, one of them help us build a bridge between two important approaches of the genetic code, the one where the amino acids are taken in their electrically neutral state (shCherbak's approach) and the other where these are considered in their real physiological state with some of them charged (Downes-Richardson's approach). The above algorithms, poetically called by mathematicians "mathemagical black holes", will be described and used in section 3.

## 2. THE TWO SIDES OF THE SAME "COIN"

### 2.1 "Head": The genetic code in one number

Using the head/tail-metaphor explained in the introduction, we first look at the head. Here, we have, *given*, the number 23!

$$23!=25852016738884976640000 \qquad (1)$$

It is written in the common place-value Hindu-Arabic decimal system notation. As a mathematical object, it also appears to be a *Gödel Number* (see the tail-sub-section below) and could describe, as we shall see below, the multiplet structure of the genetic code and its degeneracy structure as well. This would not be so odd to proceed this way since for example theorists using group theory to study the genetic code first look for a "good" irreducible representation of dimension 64 of some "good" candidate group chosen from several possible groups, as a starting point. Also, an irreducible representation is nothing but a matrix with numbers as entries. They first encode all their objects (the codons) in *one* matrix (of dimension 64) for latter manipulation such as symmetry-breaking and so on. In our approach, we encode our objects (20 amino acids and 3 stop signals) in *one* number of "length" 23, which could successfully not only encode the twenty amino acids and the three stop signals but could also contain a mean to compute the degeneracies in a very nice way. The 23 decimal digits in equ.(1) could be sorted to get the following *amino-acid-digit* assignment pattern (see the assignment details in Négadi, 2007)

$$\begin{array}{c} 5\text{ "quartets": } \{3, 5, 5, 7, 7\} \\ 3\text{ "sextets": } \{1, 2, 9\} \\ 9\text{ "doublets": } \{4, 4, 6, 6, 6, 8, 8, 8, 8\} \\ 1\text{ "triplet": } \{2\} \\ 2\text{ "singlets": } \{0, 0\} \\ 3\text{ "stops": } \{0, 0, 0\} \end{array} \qquad (2)$$

and this latter proves to be perfectly capable to fit the following well known multiplet structure of the genetic code[1]:

---

[1] The one-letter code for the amino acids is used: Glycine G, Alanine A, Proline P, Threonine T, Cysteine C, Asparagine N, Aspartic acid D, Lysine K, Glutamine Q, Glutamic acid E, Histidine H, Phenylalanine F, Tyrosine Y, Isoleucine I, Methionine M, Tryptophane W.

$$\text{5 quartets: \{G, A, P, V, T\}}$$
$$\text{3 sextets: \{S, L, R\}}$$
$$\text{9 doublets: \{C, N, D, K, Q, E, H, F, Y\}} \quad (3)$$
$$\text{1 triplet: \{I\}}$$
$$\text{2 singlets: \{M, W\}}$$
$$\text{3 stops: \{UAA, UAG, UGA\}}$$

Note that the only "multiplets" with non-prime digits are the "sextets" and the "doublets" and this property will be instrumental in the computation of the degeneracy. Of course, there is a residual permutational freedom inside each multiplet but this is absolutely harmless as our results involve *only sums* (of the numbers) inside them, in Equs.(2) and (2)'. This situation is much the same when using group theory: there remains a last permutational freedom in the assignments for the amino acids (see for example Hornos et al. 2001). Now we compute the degeneracy of the five multiplets from the digits in Equ.(2) in the following way. Call $\nu_i$ the number of digits in a given multiplet i (=1, 2, 3, 4, 6) and $\sigma_i$ the sum of the *prime factors* of the numbers, or digits, in the multiplet i, shown in Equ.(2)' below, and *without repetition*[2] (i.e., counting the original digits from Equ.(2) only one time)

$$\text{"quartets": \{3, 5, 5, 7, 7\}}$$
$$\text{"sextets": \{1, 2, 3} \times \text{3\}}$$
$$\text{"doublets": \{2} \times \text{2, 2} \times \text{2, 2} \times \text{3, 2} \times \text{3, 2} \times \text{3, 2} \times \text{2} \times \text{2, 2} \times \text{2} \times \text{2, 2} \times \text{2} \times \text{2, 2} \times \text{2} \times \text{2\}} \quad (2)'$$
$$\text{"triplet": \{2\}}$$
$$\text{"singlets": \{0, 0\}}$$
$$\text{"stops": \{0, 0, 0\}}$$

This is the same as taking the sum of the $a_0(n)$-functions of the digits, mentioned in the introduction. As a result, we have

$$\text{"quartets": } \nu_4 = 5, \; \sigma_4 = 3+5+7 = 3 \times 5 = 15$$
$$\text{"sextets": } \nu_6 = 3, \; \sigma_6 = 1+2+(3+3) = 3 \times 3 = 9 \quad (M_1)$$

$$\text{"doublets": } \nu_2 = 9, \; \sigma_2 = (2+2)+(2+3)+(2+2+2) = 9+6 = 15$$
$$\text{"triplet": } \nu_3 = 1, \; \sigma_3 = 2 \quad (M_2)$$
$$\text{"singlets": } \nu_1 = 2, \; \sigma_1 = 0$$
$$\text{"stops": } \nu_{stop} = 3, \; \sigma_{stop} = 0$$

Equs.($M_1$) *and* ($M_2$) are nothing but the *exact* Rumer's division components $M_1$ and $M_2$ of the genetic code table (Rumer, 1966). In $M_1$, also called group-IV, there are 8 quartets (32 codons) and each quartet codes for the same amino acid. In $M_2$ (group-I, group II, group-III and 3 stops; 32 total codons) this is not the case. Equivalently in $M_1$ the third base in the codon does not need to be specified in order to define an amino acid. In $M_2$ three bases have to be specified in order to define an amino acid or a stop signal. Here $\nu_i$ is identified with the number of amino acids in the multiplet i and $\sigma_i$ is associated to the total degeneracy of the multiplet. We have therefore that the total number of codons in a given multiplet i is given by $\nu_i+\sigma_i$ for all cases except for i=2 and 6 where, fortunately and interestingly, the sextets and the doublets are encoded in a subtle (and correct) way as we now see. In $M_1$ there are 5 quartets and 15 degenerate codons (15=5×3), with 20 total codons (=5+15), and 3 quartet-parts of the 3 sextets $S^{IV}$, $L^{IV}$ and $R^{IV}$ with 9 degenerate codons (3×3) and 12 total codons. In $M_2$ there are 9 doublets and 15 (=9+6) degenerate codons, 1 triplet with 2 degenerate codons, 2 singlets (degeneracy zero) and finally 3 stops also with zero σ-contribution. Note importantly that the σ-function for the nine doublets encodes, in a *taylor-made* manner, not only their 9 required degenerate codons, as the term 2+2+2+3, but also the 3 doublet-parts $S^{II}$, $L^{II}$ and $R^{II}$ of the three sextets with degeneracy 6, as the term (2+2+2). Moreover the above 9 degenerate codons correspond exactly to 3 in $U^{(1)}$ (F, Y, C), 2 in $C^{(1)}$ (H, Q), 2 in $A^{(1)}$ (N, K) and finally 2 in $G^{(1)}$ (D, E), see Table 1. In summary, there are 8 amino acids and 24 degenerate codons in $M_1$ (Equ.($M_1$)) and 12 amino acids, 17 degenerate codons and 3 stops in $M_2$. (Equ.($M_2$)). The general pattern for "$M_1+M_2$" (all meaningful coding codons for amino acids) is

$$32+29=61 \qquad (4)$$

---

[2] Taking the sum of the prime factors *without* repetition is equivalent to *discarding* the multiplicity of the digits in Equ.(2). In fact, it applies only to the quartets and the doublets as there is no multiplicity in the others.

Now to make contact with the usual picture of the five individual multiplets and in view of the above mathematical structure, we could (legitimately) *borrow* the three doublet-parts of the sextets, i.e., the term 2+2+2=6, from the $\sigma_2$-contribution in the benefit of the $\sigma_6$-part:

$$\sigma_2=9+6,\ \sigma_6=9 \rightarrow \sigma_2+\sigma_6=(9+6)+9 \rightarrow \sigma_2'=9,\ \sigma_6'=(6+9)=15 \qquad (5)$$

Formally it appears as the "exchange" $\sigma_2 \leftrightarrow \sigma_6$. The 3 doublet-parts $S^{II}$, $L^{II}$ and $R^{II}$, as doublets of codons, are subject to the FFMcG-correspondence (Findley, Findley and MacGlynn, 1982) which states that there exists a *one-to-one correspondence from one member of a doubly degenerate codon pair to the other member*. This correspondence was already used (Négadi, 2009) for the 9 doublets, in connection with the carbon atom content in the 23 AASs. Let us now convert the term 2+2+2 of the 3 doublet-parts, using the FFMcG-correspondence, as follows

$$2+2+2 \xrightarrow{\text{"FFMcG"}} (1+1+1)+(1+1+1) \leftrightarrow 3+3 \qquad (6)$$

The "borrowing" process mentioned above leads to two possible "pictures" (i)

$$\begin{aligned} \nu_4 &= 5, \; \sigma_4 = 15 \\ \nu_6 &= 3, \; \sigma_6' = (9+3+3) = 15 \\ \nu_2 &= 9, \; \sigma_2' = 9 \\ \nu_3 &= 1, \; \sigma_3 = 2 \\ \nu_1 &= 2, \; \sigma_1 = 0 \\ \nu_{stop} &= 3, \; \sigma_{stop} = 0 \end{aligned} \qquad (7)$$

and (ii)

$$\begin{aligned} \nu_4 &= 5, \; \sigma_4 = 15 \\ \nu_6' &= 3+3, \; \sigma_6' = (9+3) = 12 \\ \nu_2 &= 9, \; \sigma_2' = 9 \\ \nu_3 &= 1, \; \sigma_3 = 2 \\ \nu_1 &= 2, \; \sigma_1 = 0 \\ \nu_{stop} &= 3, \; \sigma_{stop} = 0 \end{aligned} \qquad (8)$$

using Equ.(5). In case (i) we have the usual pattern {5, 3, 9, 1, 2} with 20 amino acids, 41 degenerate codons (15+15+9+2) and 3 stops: 20+41+3=64. In case (ii), we have the other picture where the pairs $X^{II/IV}$ (X=S, L, R, see above) are considered as different objects (Négadi, 2008) which picture is linked to the degeneracy at the first (codon)-base position. Here we have 23 objects called by us amino acids signals, AAS (Négadi, 2008), the 17 usual amino acids, other than the 3 sextets, to which we add the six objects $\{S^{II}, S^{IV}, L^{II}, L^{IV}, R^{II}, R^{IV}\}$. In this case the total degeneracy is equal to 38 (15+12+9+2) and the general pattern (ignoring for simplicity the 3 stops) is

$$23+38=61 \qquad (9)$$

2.2 "Tail": The genetic code from Gödel encoding

Let us now consider the other side of the "coin" and take the prime-factorization of 23! in Equ.(1)

$$23! = 2^{19} \times 3^9 \times 5^4 \times 7^3 \times 11^2 \times 13 \times 17 \times 19 \times 23 \qquad (10)$$

One thing that could be seen at once is the sum of the exponents, 41, which appears to be equal to the total number of degenerate codons in the case of the standard genetic code (61-20). In fact we have much more. The expression in Equ.(10) is identical with what in mathematics is called a *Gödel encoding* of a number sequence. More precisely Equ.(10) is the *Gödel Number* associated to the (Gödel) encoding of the following sequence of integers (the exponents)

$$S = [19; 9, 4, 3, 2, 1, 1, 1, 1] \qquad (11)$$

The remarkable fact is that the above sequence gives a faithful *inventory*, or description, of the degeneracies when splitting the genetic code table (see Table 1) into two equal moieties and this in two different cases (i) $U^{(1)}+C^{(1)}$ and $A^{(1)}+G^{(1)}$ the so-called "first-base-pyrimidine/purine series" on the one hand, and (ii) $U^{(1)}+G^{(1)}$ and $A^{(1)}+C^{(1)}$, the "first-base-keto/amino series", on the other. $X^{(1)}$ with X=U, C, A and G is the set of all 16 codons with first-base X.

|   | $U^{(1)}$ |   |   |   | $C^{(1)}$ |   |   |
|---|---|---|---|---|---|---|---|
| UUU F | UUC F | UCU S | UCC S | CUU L | CUC L | CCU P | CCC P |
| UUA L | UUG L | UCA S | UCG S | CUA L | CUG L | CCA P | CCG P |
| UAU Y | UAC Y | UGU C | UGC C | CAU H | CAC H | CGU R | CGC R |
| UAA stop | UAG stop | UGA stop | UGG W | CAA Q | CAG Q | CGA R | CGG R |
| AUU I | AUC I | ACU T | ACC T | GUU V | GUC V | GCU A | GCC A |
| AUA I | AUG M | ACA T | ACG T | GUA V | GUG V | GCA A | GCG A |
| AAU N | AAC N | AGU S | AGC S | GAU D | GAC D | GGU G | GGC G |
| AAA K | AAG K | AGA R | AGG R | GAA E | GAG E | GGA G | GGG G |
|   | $A^{(1)}$ |   |   |   | $G^{(1)}$ |   |   |

Table 1. The standard genetic code table ($M_1$ in dark, $M_2$ in light)

We have also shown recently (Négadi, 2009) that there exist symmetry transformations, including Rumer's transformation, leaving the sequence (11) "invariant" and therefore robust against transformations (mutations) on the first base of the codon. We summarize in the following the results.
There are 19 degenerate codons in $U^{(1)}+C^{(1)}$ and 22 degenerate codons in $A^{(1)}+G^{(1)}$, in case (i). Also there are 19 degenerate codons in $U^{(1)}+G^{(1)}$ and 22 degenerate codons in $A^{(1)}+C^{(1)}$, in case (ii). This repetition comes simply from the fact that $C^{(1)}$ and $G^{(1)}$ have the *same* degeneracy structure. Now when counting the degeneracies, in each one of the two moieties, it is quite natural to keep one of the them "hidden" and counts in the detail in the other, and repeat the operation in the other way. Applying this procedure here we begin by case (i). We could describe $U^{(1)}+C^{(1)}$ having 19 degenerate codons, collectively or $U^{(1)}+C^{(1)}$ "hidden", by the term $2^{19}$ in Eq.(3) and $A^{(1)}+G^{(1)}$ having 22 degenerate codons, in some detail, by the remaining terms describing the following groups

$$\begin{aligned} &\text{V+A+G: } 9 \\ &S^{II}+R^{II}\text{: } 4 \\ &\text{T: } 3 \\ &\text{I: } 2 \\ &\text{D+E+N+K: } 1+1+1+1 \end{aligned} \qquad (12)$$

where we indicated, in parenthesis, the degenaracies or partial degeneracies (the exponents). Conversely and considering a permutation in (3) which does not change the numerical value, we could describe $A^{(1)}+G^{(1)}$ having 22=19+3 degenerate codons, collectively, by the term $2^{19} \times 7^3$, and $U^{(1)}+C^{(1)}$ having 19 degenerate codons, in some detail, by the remaining terms describing the groups

$$\begin{aligned} &L^{IV}+P+R^{IV}\text{: } 9 \\ &S^{IV}+F\text{: } 4 \\ &L^{II}\text{: } 2 \\ &\text{H+Q+C+Y: } 1+1+1+1 \end{aligned} \qquad (13)$$

where as above we have indicated the degeneracies. In case (ii), we could describe $U^{(1)}+G^{(1)}$ having 19 degenerate codons, collectively, by the term $2^{19}$ in Eq.(3) and $A^{(1)}+C^{(1)}$ having 22 degenerate codons, in some detail, by the remaining terms describing the following groups

$$\begin{aligned} &L^{IV}+P+R^{IV}\text{: } 9 \\ &S^{II}+R^{II}\text{: } 4 \\ &\text{T: } 3 \\ &\text{I: } 2 \\ &\text{H+Q+N+K: } 1+1+1+1 \end{aligned} \qquad (14)$$

Conversely, as in case (i) and using the same permutation in (3), we could describe $A^{(1)}+C^{(1)}$ having 22=19+3 degenerate codons, collectively, by the term $2^{19} \times 7^3$, and $U^{(1)}+G^{(1)}$ having 19 degenerate codons, in some detail, by the remaining terms describing the groups

$$\begin{aligned} &\text{V+A+G: } 9 \\ &S^{IV}+F\text{: } 4 \\ &L^{II}\text{: } 2 \\ &\text{D+E+C+Y: } 1+1+1+1 \end{aligned} \qquad (15)$$

In sub-section 2.1 we started with the number 23! (the "tail") and deduced the multiplet structure, as present in the division of the genetic code table into two equal moieties, $M_1$ and $M_2$, or Rumer's division, and computed the degeneracy from the decimal digits associated to the 20 amino acids and the 3 stops. Also, we deduced the usual five multiplets in two possible "pictures" (i) and (ii). Now it is clear that one could instead begin by the 'tail' which as we have seen in this section codifies the degeneracy as present in the division of the genetic code table also into two equal moieties, the "first-base-pyrimidine/purine series" on the one hand, and the "first-base-keto/amino series", on the other. In both cases the degeneracy could be encoded as a sequence of integers (see Equ.(11) and, next, this sequence is Gödel encoded to give back the Gödel number 23!, that is the "head". *These are the two sides of the same coin*.

## 3. SOME APPLICATIONS

### 3.1 Including the 21$^{st}$ and 22$^{nd}$ amino acids Selenocysteine and Pyrrolysine

As a first application we show that our arithmetic model of the genetic code shows a certain plasticity and could incorporate the recently discovered 21$^{st}$ and 22$^{nd}$ amino acids. In addition to the twenty canonical amino acids, it is known today that there exist two other (co-translationally) *encoded* amino acids for *de novo* synthesis of proteins the "twenty first" Selenocysteine (Sec) and the "twenty second" Pyrrolysine (Pyl), see the recent study by Lobanov et al. (2006). They are respectively coded by the codons UGA (opal) and UAG (amber), which code usually for stop signals (see Table 1). It is to be noted that both use tRNAs that are first charged with standard (canonical) amino acids, respectively serine and lysine, and next converted in a second (biochemical) step to Selenocysteine and Pyrrolysine. It is easy to see that the inclusion of one of the above amino acids, or even both of them, is possible in our model. As a matter of fact, we have that the *stop signals and the* (amino acids) *singlets are both assigned the digit zero* (Négadi, 2007). Including for example Sec, as singlet, would give, from Equ.(2), the following pattern

$$
\begin{aligned}
&5 \text{ "quartets": } \{3, 5, 5, 7, 7\} \\
&3 \text{ "sextets": } \{1, 2, 9\} \\
&9 \text{ "doublets": } \{4, 4, 6, 6, 6, 8, 8, 8, 8\} \\
&1 \text{ "triplet": } \{2\} \\
&3 \text{ "singlets": } \{0, 0, 0\} \\
&2 \text{ "stops": } \{0, 0\}
\end{aligned}
\qquad (16)
$$

In terms of amino acids we have

$$\begin{aligned}&\text{5 quartets: \{G, A, P, V, T\}}\\&\text{3 sextets: \{S, L, R\}}\\&\text{9 doublets: \{C, N, D, K, Q, E, H, F, Y\}}\\&\text{1 triplet: \{I\}}\\&\text{3 singlets: \{M, W, Sec\}}\\&\text{2 stops: \{UAA, UAG\}}\end{aligned} \qquad (17)$$

In this way our model could describe the genetic code with 21 amino acids (5+3+9+1+3) and it is clear that the inclusion of Sec or/and Pyl could be analogously done for a genetic code with 21 or 22 amino acids where there are now in this latter case four singlets {M, W, Sec, Pyl} and only one stop signal UAA. In this case there are 5 quartets, 3 sextets, 9 doublets, 1 triplet and finally 4 singlets, i.e., 22 amino acids. It is interesting to note that our model shows a limit for a hypothetical existence of a 23$^{rd}$ *concomitant* amino acid (assuming that it uses the stop codon UAA) because in this case there would be 5 singlets and *no* stop signal left and this would be clearly harmful for protein synthesis. This is also the conclusion of the authors cited above in their bioinformatics research for the existence of a 23$^{rd}$ amino acid: it seems unlikely.

3.2 The degeneracy and the chemical composition are connected

Our second application in this paper concerns the derivation of several *new* atomic patterns involving carbon, hydrogen, oxygen, nitrogen and sulfur, the atomic constituents of proteins, and also atom-number patterns, all present in the genetic code table, directly from the digits in our model, in also a striking "taylor-made" manner. We are going to see in this sub-section that many interesting results could be made apparent when considering the totality of "61 amino acids" in place of just 20. This is obviously equivalent to considering all 61 meaningful codons, that is, including the degeneracy. One example of such finding (here mass balance) was published some years ago (Downes and Richardson, 2002; see also Kashkarov et al. 2002)[3]. By considering a specific choice for computing "side-chain masses" (SCM) for the amino acids and their "main-chain masses" (MCM), they showed that the difference $\Delta M = SCM - MCM$, or the deviation from mass balance, is strictly 0 for the standard genetic code when considering "61 amino acids". The passage from 20 to 61 seems therefore interesting and we shall show in the following that this is also the case[4]. Now using Table 2 we have

$$\text{180 carbon atoms} \qquad (18)$$
$$\text{358 hydrogen atoms} \qquad (19)$$

---

[3] The authors take four amino acids (D, Q, K, R) ) in their ionized form, use residues in place of molecules (one water molecule less) for the amino acids. Also, proline has 73-18=55 nucleons in its block and 42 nucleons in its side-chain.

[4] I am grateful to Vladimir shCherbak for very enlightening email discussions, concerning this subtle point, and also for sending me the computer code GeneAbacus imagined by him and written by his student A. Berlizev (see section 3.3).

$$594 \text{ atoms} \qquad (20)$$

in "61 amino acids" side-chains. In this sub-section we consider two classifications of the 20 canonical amino acids. The first corresponds to the grouping of the amino acids into two sets constituted by the quartets and the sextets, on the one hand, and the doublets, the triplet, and the singlets, on the other. We call it Petoukhov's classification as it has been studied in great detail by Petoukhov (Petoukhov, 2001). There are two sets, the so-called 8 "vowel" amino acids [G, P, S, L, A, R, V, T] and the "12" consonant amino acids [C, N, Q, E, I, W, H, D, K, F, M, Y]:

| Vowel aas | | Consonant aas | | |
|---|---|---|---|---|
| G | A | C | I | K |
| P | R | N | W | F |
| S | V | Q | H | M |
| L | T | E | D | Y |
| 38 codons | | 23 codons | | |

Note that the number of codons in each one of the two sets is in agreement with the pattern in Equ.(9) but of course with a different interpretation. This classification is a good starting point to establish the existence of *balances* in carbon atom numbers and also in nitrogen-oxygen-sulfur (NOS) numbers.

| M | aa | H | C | N/O/S | Atom # | Nucleon # |
|---|---|---|---|---|---|---|
| 4 | P | 5 | 3 | 0 | 8 | 41 |
| | A | 3 | 1 | 0 | 4 | 15 |
| | T | 5 | 2 | 0/1/0 | 8 | 45 |
| | V | 7 | 3 | 0 | 10 | 43 |
| 6 | G | 1 | 0 | 0 | 1 | 1 |
| | S | 3 | 1 | 0/1/0 | 5 | 31 |
| | L | 9 | 4 | 0 | 13 | 57 |
| | R | 10 | 4 | 3/0/0 | 17 | 100 |
| | F | 7 | 7 | 0 | 14 | 91 |
| | Y | 7 | 7 | 0/1/0 | 15 | 107 |
| | C | 3 | 1 | 0/0/1 | 5 | 47 |
| 2 | H | 5 | 4 | 2/0/0 | 11 | 81 |
| | Q | 6 | 3 | 1/1/0 | 11 | 72 |
| | N | 4 | 2 | 1/1/0 | 8 | 58 |
| | K | 10 | 4 | 1/0/0 | 15 | 72 |
| | D | 3 | 2 | 0/2/0 | 7 | 59 |
| | E | 5 | 3 | 0/2/0 | 10 | 73 |
| 3 | I | 9 | 4 | 0 | 13 | 57 |
| 1 | M | 7 | 3 | 0/0/1 | 11 | 75 |
| | W | 8 | 9 | 1/0/0 | 18 | 130 |
| Total | | 117 | 67 | 9/9/2=20 | 204 | 1255 |

Table 2.
The atomic composition of the 20 amino acids
(H: hydrogen #; C: carbon #; NOS: nitrogen-oxygen-sulfur #; Atom: atom #; nucleon: nucleon #)

To reveal these balances, one must go to "61 amino acids" as mentioned above, including in this way the degeneracy. As a result we have 90 carbon atoms in the quartets and the sextets (9×4+9×6) and 90 carbon atoms in the doublets, the triplet, and the singlets (33×2+4×3+3+9). Also we have 28 NOS-atoms in the quartets and the sextets (1×4+4×6) and 28 NOS-atoms in the doublets, the triplet, and the singlets (13×2+0+1+1). We have therefore two *perfect balances* which have apparently never been mentioned. In the above classification the patterns for carbon, hydrogen and atom numbers are as follows

$$\text{Carbon: } 90+90=180 \tag{21}$$
$$\text{NOS: } 28+28=56 \tag{21'}$$
$$\text{Hydrogen: } 216+142=358 \tag{22}$$
$$\text{Atom number: } 334+260=594 \tag{23}$$

Below, we shall show how these two last patterns for hydrogen and atom numbers arise quite naturally from certain mathematical consequences of our genetic code defining number 23!. The second classification considered is Račocević's Cyclic Invariant Periodic System or CIPS (Račocević, 2009). It is visualized below in Table 3

| P | aa | C | H | At. | aa | C | H | At. |
|---|----|----|----|-----|----|----|----|-----|
| 5 | F | 7 | 7 | 14 | Y | 7 | 7 | 15 |
| 4 | L | 4 | 9 | 13 | A | 1 | 3 | 4 |
| 3 | Q | 3 | 6 | 11 | N | 2 | 4 | 8 |
| 2 | P | 3 | 5 | 8 | I | 4 | 9 | 13 |
| 1 | T | 2 | 5 | 8 | M | 3 | 7 | 11 |
| 1 | S | 1 | 3 | 5 | C | 1 | 3 | 5 |
| 2 | G | 0 | 1 | 1 | V | 3 | 7 | 10 |
| 3 | D | 2 | 3 | 7 | E | 3 | 5 | 10 |
| 4 | K | 4 | 10 | 15 | R | 4 | 10 | 17 |
| 5 | H | 4 | 5 | 11 | W | 9 | 8 | 18 |

Table 3. Račocević's Cyclic Invariant Periodic System

Račocević's explains that "the positions of the amino acids follow from their strict physical and chemical properties and also from a pure formal determination by the golden mean" (see his very recent paper, 2009). In this classification there are five classes of amino acids numbered from 1 through 5, gathered into two superclasses, the *primary superclass* PSC (even classes 2 and 4) shown here in *light*, and the *secondary superclass* SSC (odd classes 1, 3 and 5), shown in *dark*. P in the Table is for Position and is in fact the class number. In the CIPS classification the carbon composition, considering only the 20 amino acids is given by PSC: 23, SSC: 44 and this pattern, 23+44=67, could also be deduced from our formalism by considering the second picture of the amino acids (see sub-section 2.1 and below). Now, we return to hydrogen and atom numbers in "61" amino acids. In the CIPS they are as follows

$$\text{Hydrogen number: PSC: 225, SSC: 133} \rightarrow 225+133=358 \quad (24)$$

$$\text{Atom number: PSC: 341, SSC: 253} \rightarrow 341+253=594 \quad (25)$$

In the following we shall also make *contact* between these patterns and our arithmetic model based on the number 23! but before let us recall another result about hydrogen. There are 117 hydrogen atoms in 20 amino acids, see Table 2 (assuming shCherbak's choice, i.e., proline has 41 nucleons in its side-chain). For the "41" others, corresponding to the 41 degenerate codons, there are 241 hydrogen atoms with a total of 358. Now, from Equs.(1) or (2), the sum of the number of all digits, 18 (zeros do not contribute), and their total sum, 99 is equal to 117. Also, from Equ.(11) using the functions $a_0$ and $\Omega$ (see the introduction), we have $a_0(23!)+\Omega(23!)=241$ so that 117+241=358, *exactly* as described above in terms of hydrogen atom numbers. The two parts correspond perfectly: 117 is from the head, Equ.(1), which describes the 20 amino acids and 241 is from the tail, Equ.(11), which describes the degeneracy (41 codons).

*15*

Note that the number 117 consituates something like an "invariant" as it could be derived another way, this time not from Equ.(2) but from Equ.(2)' where the digits themselves are written in prime-factorization form. The new number of factors rises from 18 to 32 while their total sum falls down from 99 to 85 for all the five multiplets but the grand sum (number of factors-and-their numbers) remains constant: 18+99=117 → 32+85=117, hence the term "invariant". Let us now make our promised *contact*. As a first gush of results, take for example the sum in Equ.(24) and introduce 0=−(2+2+5)+(2+2+5) where the numbers in the parenthesis is the sum of the digits of the total number of hydrogen atoms in Račocević's *primary superclass* PSC. We have

$$225-(2+2+5)+(2+2+5)+133=216+142 \quad (26)$$

to be compared with Equ.(22), from the first (Petoukhov's) classification above. Also, introducing now 0=+(2+2+5+1+3+3)−(2+2+5+1+3+3), still in Equ.(24) where the numbers in the parenthesis is the sum of the digits of the total number of hydrogen atoms in the whole set (PSC+SPC)

$$225+(2+2+5+1+3+3)-(2+2+5+1+3+3)+133=241+117 \quad (27)$$

Starting from Račocević's PICS we obtain the hydrogen atom pattern derived above from our relations using arithmetic functions (see above) and also the number of hydrogen atoms in 20 aas (117) and in "41" degenerate amino acids (241), as also mentioned above. Alternatively, one could introduce 0 in the second members of Equs.(26) and (27) and finds the first members. There is therefore a passage between several different approaches. Consider now the atom composition. We shall show that the link with Račocević's work is direct. Consider again the number 23! in Equ.(1) and look at it as $2\times10^{22}+5\times10^{21}+8\times10^{20}+\ldots+0\times10^{1}+0\times10^{0}$, i.e., engaging the "place-value" of our 23 digits. Note that here the zeros contribute through their place-value. The sum of the place-values (the exponents of the powers of 10) for the 23 digits is equal to 253 (0+1+2+…+22) and gives us the atom number in Račocević's *secundary superclass* SSC, see Equ.(25). As for the other part in Equ.(25), we call up also the arithmetic function $a_1$ for 23! (sum of the prime factors without multiplicity) and obtain, by taking the sum of all three arithmetic functions functions $a_0$, $a_1$ and $\Omega$, $a_0(23!)+a_1(23!)+\Omega(23!)=200+100+41=341$. This is the desired result for the atom number in Račocević's *primary superclass* PSC. *In fine*, we have that several arithmetical properties of 23! "compete" constructively to produce the exact atom number pattern shown by Račocević's Cyclic Invariant Periodic System, when this latter is split into two *superclasses*, that is 253+341=594. Also, and interestingly, as for the hydrogen number 358 (see above), here the two parts also correspond perfectly: 253 is from the head, Equ.(1), which describes the 20 amino acids and 341 is from the tail, Equ.(11), which describes the degeneracy. The link between our arithmetic model and Račocević's PICS is manifestly *direct*. Finally, we end this sub-section by a second gush of results, obtained not exactly from 23! itself but from certain mathematical consequences of it we now explain. It is related to the mathematically interesting number 123. The second picture, (ii), described in Equ.(8) leads to this number. As a matter of fact we have 17 amino acids not "degenerate" at the first-base position and 3 "doubly-degenerate" AAS ($S^{II}$, $S^{IV}$, $L^{II}$, $L^{IV}$, $R^{II}$, $R^{IV}$) so that by assigning the numbers 1 through17 to the former group and 18 through 23 to the latter, we obtain by taking the sums in each set 153+123. This relation was studied recently by us (Négadi, 2009) and we have shown that it could lead to 67, the carbon atom number in the 20 amino acids side-chains, but the carbon-patterns obtained from our numbers 153 and 123 were favourably compared to the genetic code carbon-pattern only modulo the mathematical "trick" (-1+1=0). Here we obtain directly the exact carbon atom pattern for Račocević's CIPS, without any trick at all. As a matter of fact we have $a_0(153)$= (17+3+3)=23 and $a_0(123)$=3+41=44 (23+44=67). These are nothing but the *exact* number of carbon atom in PSC and SSC, respectively. An interesting coincidence found in Račocević's CIPS concerns the three sextets *serine leucine* and *arginine* for which it appears that *their class number or position "P" coincides exactly with the number of carbon atoms in their side-chains* S: 1, L: 4, R: 4 (see Table 2). Using this fact we could write

$$a_0(153)+a_0(123)+ \qquad (28)$$
$$P(S)+P(L)+P(R)=(23+44)+(1+4+4)=67+9=76$$

This is also a relation that seems *taylor-made* to describe the carbon content in the 23 AASs, 67 for the 20 aas and 9 for the three sextets again (in this picture, the sextets are counted two times). In all triviality but interestingly, from the above equation, different combinations of the factors coming from the $a_0$-part and the position-part of the 3 sextets could be made to fit the carbon atom patterns in some known divisions, for example (i) the Rumer division $M_1/M_2$: (3+3+3+9)+(17+41)=18+58, (ii) Račocević's CIPS, again but this time in terms of 23 objects, the 23 AASs and for the *same* division PSC/SSC: (1+3+3+3+4+17)+(41+4)=31+45 (see Table 3) and finally (iii) the "first-base-pyrimidine/purine series" or $U^{(1)}+C^{(1)}/A^{(1)}+G^{(1)}$ division considered in sub-section 2.2 of this paper: (41+3+3)+(1+3+3+3+4+17)=47+29. Through the above treatment, the number 123 has gained some importance (together with 153, see below) and we now show that it could be further exploited. To see how, let us call upon an algorithm known as "mathemagical black hole" (Ecker, 2004) and precisely the number 123 is one such "black hole": it irresistibly swallows any natural number after a given number of (few) iterations for a certain process or algorithm. (We have just seen above that this number plays an important role in our so-called second "picture" of the amino acids and the corresponding carbon content.) The iterative algorithm mentioned above works as follows: start with *any* number and count the number of even digits and the number of odd digits. Write them down next to each other (by concatenation) following by their sum. Treat the result as a new number and continue the process. This latter is very quick even for big numbers. Now let us *apply this process to the decimal representation of 23!* in our main Equation (1). We have at the first iteration 16 even numbers, 7 odd ones and their sum 23, so that $I^{(1)}=16723$. The final result, after only three iterations, gives

$$I^{(1)}=16723$$
$$I^{(2)}=235 \qquad (29)$$
$$I^{(3)}=123$$

These three numbers have different size. In order to have numbers of the same size, here 3, we choose to modify (not randomly) the first of the above three iterations, $I^{(1)}$, by replacing the first 5-digit number by *the sum of the prime indices of its prime factors*. As $16723=7\times2389$ and knowing that 7 is the 4$^{th}$ prime and 2389 is the 355$^{th}$ prime, we obtain

$$\tilde{I}^{(1)}=359 \qquad (30)$$
$$I^{(2)}=235 \qquad (31)$$
$$I^{(3)}=123 \qquad (32)$$

This homogeneous triplet of numbers $\{\tilde{I}^{(1)}, I^{(2)}, I^{(3)}\}$, tightly bound to the preceeding one in Equ.(29) is interesting. For example, $\tilde{I}^{(1)}+I^{(2)}=359+235=594$ and this number is equal to the total number of atoms in "61 aas". Also, $I^{(2)}+I^{(3)}=235+123=358$ and this number is equal to the total number of hydrogen atoms in "61 aas". Finally, by subtracting the latter sum from the former one, we obtain $\tilde{I}^{(1)}-I^{(3)}=359-123=236$ and this is nothing but the total number of CNOS atoms in "61 aas". Note, for the atom number relation (359+235) its proximity to the real distribution of hydrogen atoms and CNOS atoms 358+236=594; it is only ±1 away. This is nice but we could also infer other interesting conclusions from this triplet of numbers. The total atom number in "61 amino acids" as mentioned above is equal to 594. The sum of the digits of the numbers in the *first* and *last* iterations is equal to $s^{16723}+s^{123}=s=25$. Introducing $-s+s=0$ in the sum $\tilde{I}^{(1)}+I^{(2)}$ we get

$$\tilde{I}^{(1)}+I^{(2)}=359-s+s+235=334+260=594 \qquad (33)$$

This is an exact atom number distribution for what we called above Petoukhov's classification. We have 334 atoms in the quartets and the sextets (31×4+35×6), on the one hand, and 260 atoms in the doublets, the triplet and the two singlets (96×2+13×3+11+18), on the other. Using the other alternative, we get

$$\tilde{I}^{(1)}+I^{(2)}=359+s-s+235=384+210=594 \qquad (34)$$

For comparison there are 384 atoms in the 17 first-base non-degenerate amino acids and 210 atoms in the six objects $X^{II/IV}$ (X=S, L, R). Now we consider

Račocević's CIPS classification. There are 341 atoms in PSC and 253 atoms in SSC, as we have seen above. The sum of the latter six digits could, themselves, be used and introduced in $\tilde{I}^{(1)}+I^{(2)}$ to write

$$359+235-(3+4+1+2+5+3)+(3+4+1+2+5+3)=341+253 \quad (35)$$

and we obtain precisely that pattern. Also we have $a_0(341)=42$ ($341=11\times31$) and $a_0(253)=34$ ($253=11\times23$) so that $a_0(341)+a_0(253)=76$, i.e., the number of carbon atoms in 23 objects (sextets counted two times), see below. Račocević's PSC/SSC division is obtained in only one step: $31+(11+11+23)=31+45=76$, see above after Equ.(28). We consider now hydrogen atom number and the second sum, from Equs.(31)-(32): $I^{(2)}+I^{(3)}=235+123=358$. Introducing $0=-s^{123}+s^{123}$ in this equation gives

$$235+s^{123}-s^{123}+123=241+117=358 \quad (36)$$

or, introducing this time $0=-s^{16723}+s^{16723}$ gives

$$235-s^{16723}+s^{16723}+123=216+142=358 \quad (37)$$

These are again the patterns for hydrogen, the first was already considered above and the second is the pattern for the quartets and the sextets (216) and for the doublets, the triplet and the singlets (142) in Equ.(22).

3.3 An application in proteomics

Finally, as another and last application in this paper, this time *outside* the genetic code itself, we show that *specific ratios* of certain arithmetic quantities associated to the amino acids classes in our model could corroborate favourably the experimental amino acids composition, in the corresponding classes, and averaged over a large set of recently analyzed proteomes covering the three phylogenetic domains of life (Tekaia and Yaramian, 2006). The idea behind this application is that some properties of the genetic code could "persist" at the genomic or proteomic level, as it has been shown (Downes and Richardson, 2002). These authors found a strict "mass balance" $\Delta M=0$ between the side-chain masses and the main-chain masses for 61 "aas" of the genetic code table, as found also by shCherbak and his collaborators, but went farther by showing that this phenomenon nearly persists ($\Delta M\sim0$) also at the proteomic level when considering a whole set of 203 representative species. In our arithmetic model, the 20 amino acids are assigned digits and the sum of the *prime factors* of all the digits in the five multiplets gives the total degeneracy, 41, see picture (i) above. Now, taking instead the sum of the digits themselves (not their prime factors) and also *without repetition* we have 47. The latter quantity might correspond to some more general "property", *including* the total degeneracy as a particular case. We shall see below that we could eventually link it with the amino acid composition. We have just seen, in this section, that the degeneracy and the chemical composition of the amino acids are connected so we could expect also a connection of that "property" with the amino acids composition itself. From the data of Tekaia and Yaramian mentioned above[5], we have computed the following percentages for the five multiplets the quartets:~31.4%, the sextets:~21.9%, the doublets:~36,6%, the triplet:~6,6% and finally the singlets:~3.45%. Let us define now the new quantity $\Sigma_0=47+(1+2)=50$, where the additional factor 1+2 concerns *only* the three odd-degeneracy amino acids and corresponds to the contribution of the singlet-part of the triplet (I) and the one of the two singlets (M and W) through their numbers. With this choice our model lead us to the following ratios between the sum of the digits in each multiplet and $\Sigma_0$: quartets $15/\Sigma_0=30\%$, sextets: $12/\Sigma_0=24\%$, doublets: $18/\Sigma_0=36\%$, triplet: $(2+1)/\Sigma_0=6\%$ and singlets: $2/\Sigma_0=4\%$. It is seen that the above additional factor, present also in the numerator of the three odd-degeneracy amino acids, is necessary in order to get non-aberrant results, as for example the two singlets which have the sum of their digits equal to 0; we would have obtained 4% for I and 0% for M and W. With these considerations the comparison between the percentages obtained above and those coming from experiment seems thus rather favourable. Strikingly, and as suggested to us by shCherbak, the results for a *single* gene, here the one for Homo sapiens glycyl-tRNA synthetase (739 aas), seems also not too unreasonable. As a matter of fact, using "GenAbacus" we find the following percentages: quartets:~30.8%, sextets:~21.4%, doublets:~39,6%, triplet:~4,9% and singlets:~3.3%. Does some property related to $\Sigma_0$ (and amino acids composition) also persists as in the case of the amino acids with mass balance? Also and strikingly, when the above "theoretical" percentages (integer numbers here) are evaluated for the "61 amino acids" and the five multiplets separated into two sets the sextets and the doublets, on the one hand, and the quartets, the triplet and the singlets, on the other, we obtain the following remarkable relation $(24\times6+36\times2)+(30\times4+6\times3+4\times1)=216+142=358$, which is nothing but the total number of hydrogen atoms in "61 amino acids", once again (see our equations (22), (26) and (37) above). The sextets and the doublets have played an important role in our model because they are the only two multiplets among the five where the prime factorization process brings something new; in the other three there are either primes (and $a_0(p)=p$) or zero. It is this very latter fact that led us to the correct degeneracy in section 2.1.

## 4. CONCLUDING REMARKS

---

[5] Fredj Tekaia is warmly acknowledged for sending me the data.

Let us close this paper by making some remarks. The first one concerns the link between the atomic content of the special "imino" acid proline and the number 358, which is also the total number of hydrogen atoms in "61 amino acids" is 358. This number is important because it hides proline's singularity, as we have shown in our recent work (Négadi, 2008); it describes famous shCherbak's imaginary "borrowing" of *one* nucleon (hydrogen) from proline's side-chain (42-1) in favor of its block (73+1), which borrowing paves the way for the existence of numerous beautiful nucleon-number balances or the "Pythagorean triple" (shCherbak, 2003). We have recently also shown that this number (358) could be derived from our arithmetic model in a nice way (Négadi, 2009). Observe in Table 3 (Račocević's CIPS) that, for proline, we have the position number P(Pro)=2, the carbon number: c(Pro)=3, the hydrogen number h(Pro)=5 and the sum of these last two numbers c(Pro)+h(Pro)=a(Pro)=8 which happens to be also the number of atoms a (no NOS atoms). These numbers, 2, 3, 5 and 8, are four consecutive members of the Fibonacci series. It is also striking that the number 358 hiding "proline's" singularity, in the form PI(2)+PI(179)=1+41=42 where 2 and 179 are the prime factors of 358 and PI is for Prime-Index, could be so easily obtained by simple *concatenation* of the three quantities c, h and a=c+h defined above (we drop the argument "Pro" in the following for clarity):

$$ch(c+h)=cha=c\times 10^2+h\times 10^1+a\times 10^0=358 \qquad (38)$$

as decimal place-value representation. Also, by introducing the nucleon numbers with (without) blocks and the number of atoms also with (without) blocks

$$N(n): 115(41); A(a): 17(8) \qquad (39)$$

we obtain by adding these four numbers N+n+A+a=181 which is nothing but the sum of the prime factors of 358 (2+179) and 181=$a_0$(358). Considering the sum of the $a_0$-functions of the differences $a_0$(N−n)+$a_0$(A−a)=45 and taking the whole sum we get 181+45=226 which is equal to the hydrogen content in $A^{(1)}+C^{(1)}$. The other part is given by cha-226=132, using Equ.(38), and this is the hydrogen content in $U^{(1)}+G^{(1)}$. Note also that N+A=132 and cha-132=226. Finally, we consider $a_0$(N)=28 and form $a_0$(N)+N+A+a=168 which is the hydrogen content of $A^{(1)}+G^{(1)}$ and by subtracting this number from 358, we obtain 190, the hydrogen content in $U^{(1)}+C^{(1)}$. We note finally that there are 226 hydrogen atoms in the 17 amino acids "non-degenerate" at the first-base position and 132 hydrogen atoms in the 3 "doubly-degenerate" AAS (see above and Table 2). It is quite striking but nevertheless interesting that proline contain such an amount of information. Our second remark concerns the number 359 but this time in its relation with our famous "black hole" number 123 (see below, once more about it). We could write the sum of their digits to obtain 3+5+9+(1+2+3)=23. Rearranging we get 5+9+1+2+(3+3) which could be read as 5 quartets, 9 doublets, 1 triplet, 2 singlets and finally the 6=(3+3) objects $X^{II/IV}$ (X=S, L, R), see sub-section 2.1. This is the pattern for 23 AASs. Moreover, the only non-prime digit (except 1) is 9 so that by taking its $a_0$-function $a_0(9)=3+3$ we could now write, gathering the three's, 5+(3+3+3)+1+2+3=5+9+1+2+3=20 and "9" *reappears*. This is the usual pattern for 20 aas. Applying again the operation to the non-prime number nine in the last relation gives 5+9+1+2=17 and these are our 17 first-codon position "non-degenerate" amino acids mentioned above. This remark and the preceding one together brings us to the last and most important one that is the link between what we called in the introduction shCherbak's and Downes-Richardson's approaches. In both, remarkable mass (nucleon number) balances were detected in the genetic code table. shCherbak (shCherbak, 2003) has shown, by considering the nucleons numbers (or integer molecular mass) in the set of 23 AASs, the existence of remarkable mathematical patterns, as for example the first Pythagorean triple {3,4,5} in $M_1$ and a nucleon number balance between the nucleon numbers in the side-chains (1110 nucleons) and those in the blocks (1110 nucleons) in $M_2$. Also, he showed another nucleon number balance in $U^{(1)}+C^{(1)}$ and many other remarkable arithmetic identities throughout his paper. From the reading of his paper, we have also found, some years ago (Négadi, 2001), that the distribution pattern of the nucleon numbers in $A^{(1)}+G^{(1)}$, the complement of $U^{(1)}+C^{(1)}$, codifies the degeneracy in $M_1$ and $M_2$ in the form 24+17=41 (see the text above Equ.(4)). Note importantly that shCherbak worked *also* with the amino acids in their *electrically neutral* form *all* having a block with 74 nucleons and the unique and special (imino) acid *proline* having 41 nucleons in its "side-chain" (one nucleon has been borrowed from the side chain to the block). In the first approach by shCherbak there are 1443 nucleons in the 23 AASs *side-chains*, 333 in $M_1$ and 1110 in $M_2$. In the second, Downes and Richardson but also shCherbak and his collaborators (Kashkarov et al., 2002), on the other hand, considered some of the amino acids in their (physiological or "real-life") ionized form[6], all having a "block" with 56 nucleons except the unique and special (imino) acid *proline* with 55 nucleons in its "block" and 42 nucleons in its "side-chain" (in this case one water molecule, mw: 18, is removed). In the approach by the above authors one considers all "61 amino acids" and computes the "side-chain masses" (SCM) and finds 3412 nucleons and the "main-chain masses" (MCM) and finds also 3412 nucleons. There is therefore also in this case a nice nucleon number balance, or SCM=MCM, but the above authors consider instead the difference ΔM=SCM−MCM, or the deviation from mass balance, which appears to be strictly 0 for the *standard* genetic code. The link between the two approaches could be made by comparing them at the "61 amino acids" level: in the "neutral case" one has 3404 nucleons in the "61 side-chains" (see Négadi, 2009) while in the "physiological" case one has, as mentioned above, 3412 nucleons in the "61 side-chains". The relation between these two numbers, 3404 and 3412, is easy to see. In $M_2$ the *charges* are $\delta(D)=-1$, $\delta(K)=+1$, $\delta(E)=-1$ and $\delta(R^{II})=+1$ so that they compensate each other and the result is 0 while in $M_1$ one has eight more nucleons, 4 for proline ($1\times4$) and 4 for $R^{IV}$ ($+1\times4$). We have therefore 3404+8=3412. Precisely, it is our aim to "derive" below this latter interesting relation. We have already seen that the numbers 123 and 153, *both* mathematical "black holes", for two different processes, the former for any natural number and the latter for any number multiple of three, are associated to our second "picture", see above. The latter could be linked to the 17 first-base "non-degenerate" amino acids set and the former to the 3 "doubly-degenerate" AAS (S, L, R). This association is very interesting because these numbers bear something like *self-information*. Take for example the number 153. We have 153/(1+5+3)=17 which is precisely the number of first-base "non-degenerate" amino acids. That the number 123 is linked to the sextets could be seen as follows. First we begin by taking into account the "doubly-degenerate" nature of the sextets (see section 3) and we "double" it to get two copies $123_1$ and $123_2$. This gives us 6 digits (=3+3) and the sum of all 6 digis is 12 (=6+6) so that the total is equal to $3\times6=18$ which is precisely the total number of codons in the three sextets serine, leucine and arginine. This said, we next apply *another* very simple mathematical "black hole" process, like the one used above, to our "sextets-number" number 123. Remind that it is itself a "black hole" as seen above but for another process. As this latter is multiple of three (=3×41) the "black hole" in this case is the number 153. (Fancily we shall have a case of mathematical black hole "cannibalism": 153 "swallows" 123.) The algorithm works as follows: start from a number (multiple of 3), take the sum of the cubes of its digits, and restart again the iterative process till reaching the black hole 153. We get therefore the *double* (*identical*) series [123, 36, 243, 99, 1458, 702, 351, 153, 153]$_1$ and [123, 36, 243, 99, 1458, 702, 351, 153, 153]$_2$ for $123_1$ and $123_2$. In both there are 7 steps from 123 to the *first occurrence* of 153 (eight numbers). For each one of them the total sum $S$ of (i) the sum of *all* the numbers in the series, (ii) the sum of the digits in the *eight* distinct numbers mentioned above (123-153) and (iii) the number of steps from 123 to the *first* 153 gives 3412 (3318+87+7) and we have

$$123_1 \to S_1=3412 \qquad (40)$$
$$123_2 \to S_2=3412 \qquad (41)$$

---

[6] These are aspartic acid D (59 →58; δ=−1), lysine K (72 →73; δ=+1), glutamic acid E (73 →72; δ=−1) and arginine R (100 →101; δ=+1). The first numbers are the old (neutral) integer molecular weights, the second ones are the new integer (ionized) molecular weights and finally the third ones, δ, the difference between the first and the second numbers is the *charge*.

The last number in each one of the series, 153, appears two times one as a normal iteration and the second as a "check" iteration for the black hole algorithm; we therefore include it in the whole sum, (i), but discard it when considering the 7 steps from 123 to the first occurrence of 153, in (ii) and (iii). Making now the following identifications $S_1=3412 \to$ MCM and $S_2=3412 \to$ SCM we have immediately the identity $S_2 - S_1 =$ SCM−MCM$=\Delta M = 0$. Moreover taking only one of them, for the Side-Chain Mass here, we write the above total sum as (3318+87)+7=3405+7. Note that shCherbak's "view" with 3404 nucleons (see above) is not very far. By introducing −1+1=0 in the large part, we could write for example 3404+(1+7)=3404+8=3412. Remember that we have two times $\delta=\pm 1$ in $M_2$ (see above for the transformation from neutral to charged amino acids and footnote 6) so that we could (virtually) use either of them in the form −1+1=0. In the relation above, the "1" is for the "famous" one nucleon of the side-chain of proline, to be borrowed by its block, (see shChebak, 2003) and 7 is the sum of 3 other nucleons for the remaining 3 degenerate codons of proline and 4 nucleons for the quartet-part of the (positively charged) arginine: $R^{IV}$ (+1×4), see above. This is the passage formula between the two approaches, promised above.